\documentclass[epj]{svjour}
\usepackage{graphics}
\newcommand{\beq}{\begin{equation}}
\newcommand{\eeq}{\end{equation}}
\newcommand{\bea}{\begin{eqnarray}}
\newcommand{\eea}{\end{eqnarray}}
\newcommand{\bef}{\begin{figure}}
\newcommand{\eef}{\end{figure}}
\newcommand{\bce}{\begin{center}}
\newcommand{\ece}{\end{center}}
\newcommand{\eg}{{\it e.g.}}
\newcommand{\ie}{{\it i.e.}}
\newcommand{\etal}{{\it et al.}}
\def\lsim{\mathrel{\rlap{\lower4pt\hbox{\hskip1pt$\sim$}}
    \raise1pt\hbox{$<$}}}         
\def\gsim{\mathrel{\rlap{\lower4pt\hbox{\hskip1pt$\sim$}}
    \raise1pt\hbox{$>$}}}         
\begin{document}
\title{Quark Coalescence and Charm(onium) in QGP}
\author{Ralf Rapp
}                     
%
%
\institute{Cyclotron Institute and Physics Department, 
Texas A\&M University,
College Station, TX 77843-3366, U.S.A.}
\date{Received: date / Revised version: date}
%
\abstract{
The potential of heavy quarks as probes of the environment
produced in hadronic and heavy-ion reactions is discussed.
A key role is played by coalescence processes and/or
resonance formation which are promising candidates to
provide a comprehensive understanding of phenomena associated
with reinteractions of both open and hidden heavy-quark states.
} 
\maketitle
%

\section{Introduction}
\label{sec_intro}
In hadronic and heavy-ion collisions, heavy quarks ($Q$=$c$, $b$) are 
believed to be (almost) exclusively pair-produced ($Q\bar Q$) upon first 
impact in hard partonic collisions~\cite{LMV95}.  
This renders them excellent agents of the subsequently formed medium 
and their reinteractions within. The latter include: (a) coalescence
with surrounding quarks as hadronization mechanism in addition to  
fragmentation, thereby probing the chemical and kinematic properties of 
the medium~\cite{KLS81,VBH92,Hwa95,BJM02,RS03};
(b) energy loss of high-momentum $Q$-quarks~\cite{Djord05,Wied05}, 
which, with increasing interaction strength toward lower momentum, 
eventually leads to
(c) thermalization~\cite{Svet88,Must97,HR05,MT05}; and, if the latter 
can be established, 
(d) in-medium dynamics of open and hidden heavy-flavor 
states~\cite{MS86,Blasch01,RG04,GRB04,Wong04}, which is particularly 
exciting in view of recent QCD lattice 
calculations~\cite{AH04,Dat04,Umeda05} indicating the survival of 
low-lying charmonia well into the Quark-Gluon Plasma (QGP).
In this paper we will address the above issues essentially in 
that order.

\section{Coalescence in Hadronic Collisions}
\label{sec_asym}
\begin{figure}[!t]
\vspace{0.2cm}
\resizebox{0.465\textwidth}{!}{
\includegraphics{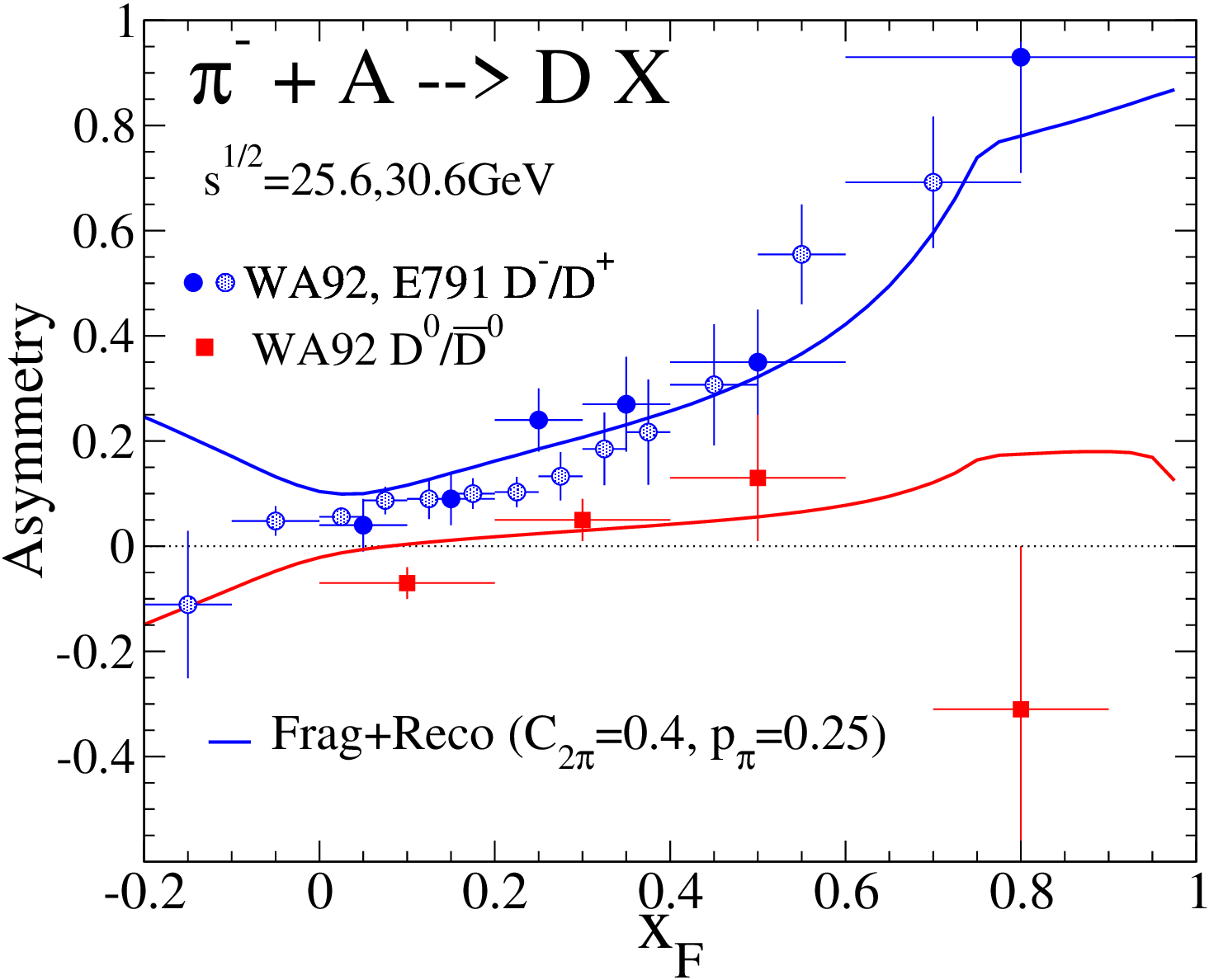}
}
\vspace{0.05cm}
\caption{$D$-meson flavor asymmetries, $A$=($N_{D_1}-N_{D_2}$)
/($N_{D_1}+N_{D_2})$, in $\pi^-$-$A$
reactions~\protect\cite{e791-96,wa92-97} compared to coalescence + 
fragmentation calculations~\protect\cite{RS03}. Upper data points and 
curves are for $D_1$=$D^-$ and $D_2$=$D^+$, whereas lower data points and 
curves are for $D_1$=$D^0$ and $D_2$=$\bar D^0$. ``Leading" particles 
($D^-$=$\bar cd$, $D^0$=$c\bar u$) are defined as sharing a valence
quark with the projectile ($\pi^-$=$d\bar u$); note that the approximate
absence of an asymmetry for $D^0$/$\bar D^0$ is accounted for in the model
due to the predominant production of $c\bar c$ pairs in the forward
direction via $\bar uu$ annihilation, rendering the valence $\bar u$
unavailable for recombination. }
\label{fig_fasym}       
\end{figure}

In elementary hadronic reactions ($pN$, $\pi N$) evidence for 
reinteractions of $c$-quarks arises from (large) flavor asymmetries in 
$D$-meson production yields. The asymmetries are most pronounced at 
forward rapidities (or $x_F$), successfully 
being attributed to coalescence of $c$-quarks with valence quarks 
of the projectile~\cite{Hwa95}. The pertinent recombination cross 
section can be written as~\cite{KLS81}
\beq
x^* \frac{d\sigma^{rec}_D}{dx_F} = \int \frac {dx_{\bar q}}{x_{\bar q}}
\int \frac{dz}{z} \ \left( x_{\bar q} z^*
\frac{d^2\sigma^{(c\bar q)}}{dx_{\bar q} dz}\right) \ 
{\cal R}(x_q,z;x_F)  \ ,
\label{coal}
\eeq
where the main elements are: (i)  the $c$-$\bar q$ production cross 
section composed of a 2-parton distribution function (2-PDF),
$f^{(2)}_{i\bar q}$ (where $i$=$g$,$q$,$\bar q$ participates in the hard
process to produce the $c\bar c$ pair), and the standard 
perturbative QCD (pQCD) $c\bar c$ cross section, and (ii) the 
$c$-$\bar q$$\to$$D$ 
recombination function, ${\cal R}$. The 2-PDF is usually factorized 
into two single PDFs with phase space correction, 
\beq
f^{(2)}_{i\bar q}= C f_{\bar q}(x_{\bar q}) \  f_i(x_i) \  
(1-x_{\bar q} -x_i)^p \ , 
\eeq
whereas ${\cal R}$ represents a $D$-meson wave function which in 
Ref.~\cite{RS03} has been assumed to be Gaussian in rapidity space,
\beq
{\cal R}(y_{\bar q},y_c,y)=
\exp(\Delta y^2/2\sigma_y^2) / \sqrt{2\pi\sigma_y^2} \ . 
\eeq
This form of the recombination function~\cite{Tak79} allows to 
generalize the coalescence formalism to include
sea-quarks~\cite{RS03}, and thus address flavor asymmetries also at 
central $x_F$, cf. Fig.~\ref{fig_fasym}.
The experimentally observed asymmetries in inclusive yields ($x_F$$>$0)
are quite appreciable, 
\eg, $D^-$/$D^+$= 1.35$\pm$0.05 (versus 1 in isospin-symmetric
fragmentation), $D^0$/$\bar{D}^0$=0.93$\pm$0.03 (vs. 1) and
$D^\pm$/($\bar{D}^0$+$D^0$)=0.415$\pm$0.01 (vs. 0.33) for fixed-target
$\pi^-N$ collisions (averaged over a weak energy dependence for
$\sqrt{s}$=19-34GeV)~\cite{wa92-97}. The data
are rather well reproduced by a combined coalescence + fragmentation 
approach~\cite{RS03} (for a somewhat different framework based
on power corrections, see Ref.~\cite{BJM02}).

\section{Open Charm in the QGP}
\label{sec_open}
Final-state interactions of heavy quarks are enhanced when embedding 
them into a heavy-ion collision, where, at ultrarelativistic energies, 
intense reinteractions of light partons are
believed to form locally thermalized matter within a  
time of $\tau$$\lsim$1fm/c. 
At high momenta $c$-quarks rescatter 
perturbatively inducing a softening of the primordially power-like 
$p_t$-spectra, with subsequent hadronization in the vacuum 
(fragmentation). The predicted suppression factors relative to $p$-$p$
collisions range from 0.2~\cite{Wied05} to 0.5~\cite{Djord05}, with
a rather small azimuthal asymmetry, $v_2$$\le$5\%~\cite{DG05}. 

\begin{figure}[!t]
\vspace{0.3cm}
\resizebox{0.48\textwidth}{!}{
\includegraphics{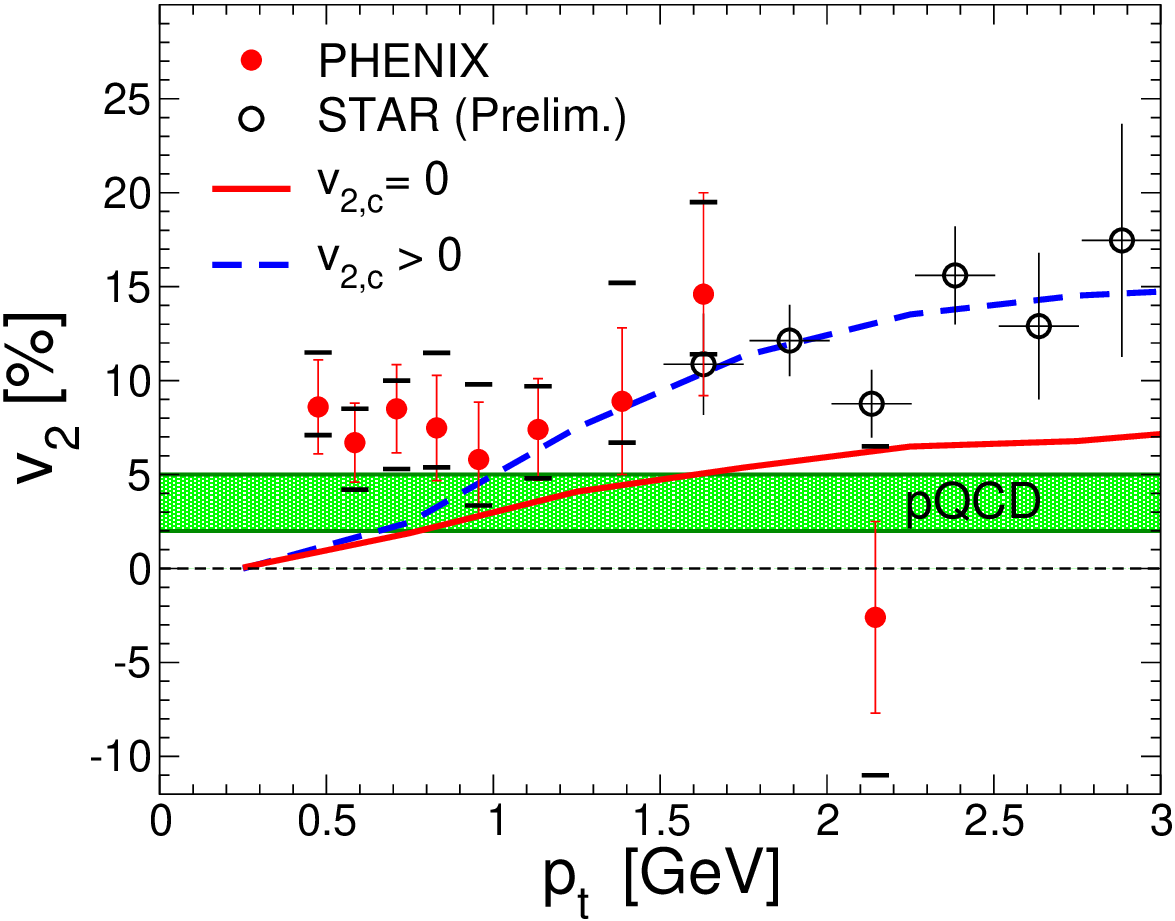}
}
\caption{Azimuthal asymmetry of ``non-photonic" single-$e^\pm$ 
spectra in minimum bias
$Au$-$Au$($\sqrt{s_{NN}}$=200GeV)~\protect\cite{phenix-v2e,star-v2e}
compared to coalescence model predictions~\protect\cite{Greco04}
using $c$-quark distributions from either $p$-$p$ collisions 
($v_2^c$=0, solid line) or assuming a transverse flow and $v_2$-profile 
as determined for light quarks from fits to light-hadron spectra 
($v_2^c$$>$0, dashed line).  The band indicates predictions from
jet-quenching~\protect\cite{DG05} applicable at sufficiently high
$p_t$.}
\label{fig_coal}
\end{figure}

Toward lower $p_t$, the phase space density of 
the medium increases and coalescence with light quarks is expected
to become competitive~\cite{Greco04,Chen05,Mol05}. The same expression, 
Eq.~(\ref{coal}), can be applied with the light quark
distributions being replaced by thermal (+ quenched pQCD) ones
as established from light hadron production systematics (also,
the recombination function ${\cal R}$ is typically substituted with
a hadron wave function in transverse momentum).  
The extension to low $p_t$ is, in principle, more controlled than for
light-light ($q$-$\bar q$) coalescence, since at the scale of the 
hadronization temperature, secondary $c$-production is negligible.  
With previously determined light-quark distributions, charmed-hadron 
spectra become a sensitive probe of the dynamics of $c$-quarks in the 
QGP. This has first been quantified in the context of ``charm-like" 
single-electron spectra in Ref.~\cite{Greco04}, showing 
that $v_2^e(p_t)$: (a) closely reflects the $v_2$ of the parent 
$D$-meson, (b) exhibits a marked difference of more than a factor
of 2 between the cases where the $c$-quark distributions are either 
taken from $p$-$p$ collisions, or assumed to follow the systematics of 
light quarks (including collective expansion), cf.~Fig.~\ref{fig_coal}. 
Current data at RHIC from PHENIX~\cite{phenix-v2e} and STAR 
(preliminary)~\cite{star-v2e} seem to favor the quasi-thermalized 
scenario. 
If confirmed, this raises at least two further questions: \\ 
(i) Is the predicted $v_2^e$ consistent with pertinent $p_t$-spectra 
(\ie, the ratio of central $Au$-$Au$ to collision-scaled $p$-$p$ 
spectra, $R_{AA}$)? \\
(ii) What are microscopic mechanisms for
thermalization of $c$-quarks (or $D$-mesons)? 

Concerning (i), it has been pointed out~\cite{Bats03} that for 
single-$e^\pm$ $p_t$-spectra in central $Au$-$Au$, $N_{coll}$-scaled 
$D$-me\-son spectra from $p$-$p$ collisions lead to results rather 
similar to a scenario based on full thermalization and collective 
flow close to hadronic freezeout ($T$$\simeq$130MeV, $v_\perp$=0.65),
due to large blue shifts with $m_D$=1.87GeV (also,
bottom-decay contributions  become significant above
$p_t^e$$\simeq$3GeV). 
However, in hydrodynamic analysis~\cite{MT05} coupled
with a Fokker-Planck treatment of $c$-quarks in the QGP, a $v_2^c$ of 
$\sim$10-15\% is associated with $R_{AA}^c$($p_t$$\ge$3GeV)$\le$0.1.   
Coalescence model calculations~\cite{Greco04}, 
based on recombination at the phase boundary, imply a suppression 
factor similar to jet quenching, $R_{AA}^D$($p_t$$\simeq$3GeV)=0.2-0.5,
but with $v_2^D$($p_t$$\simeq$3GeV)$\simeq$15\%.     
Note that when starting from $c$-quark spectra, fragmentation leads 
to a degradation, whereas coalescence to an 
increase, of the resulting $D$-meson $p_t$.  

\begin{figure}[tb!]
\vspace{0.4cm}
\resizebox{0.45\textwidth}{!}{
\includegraphics{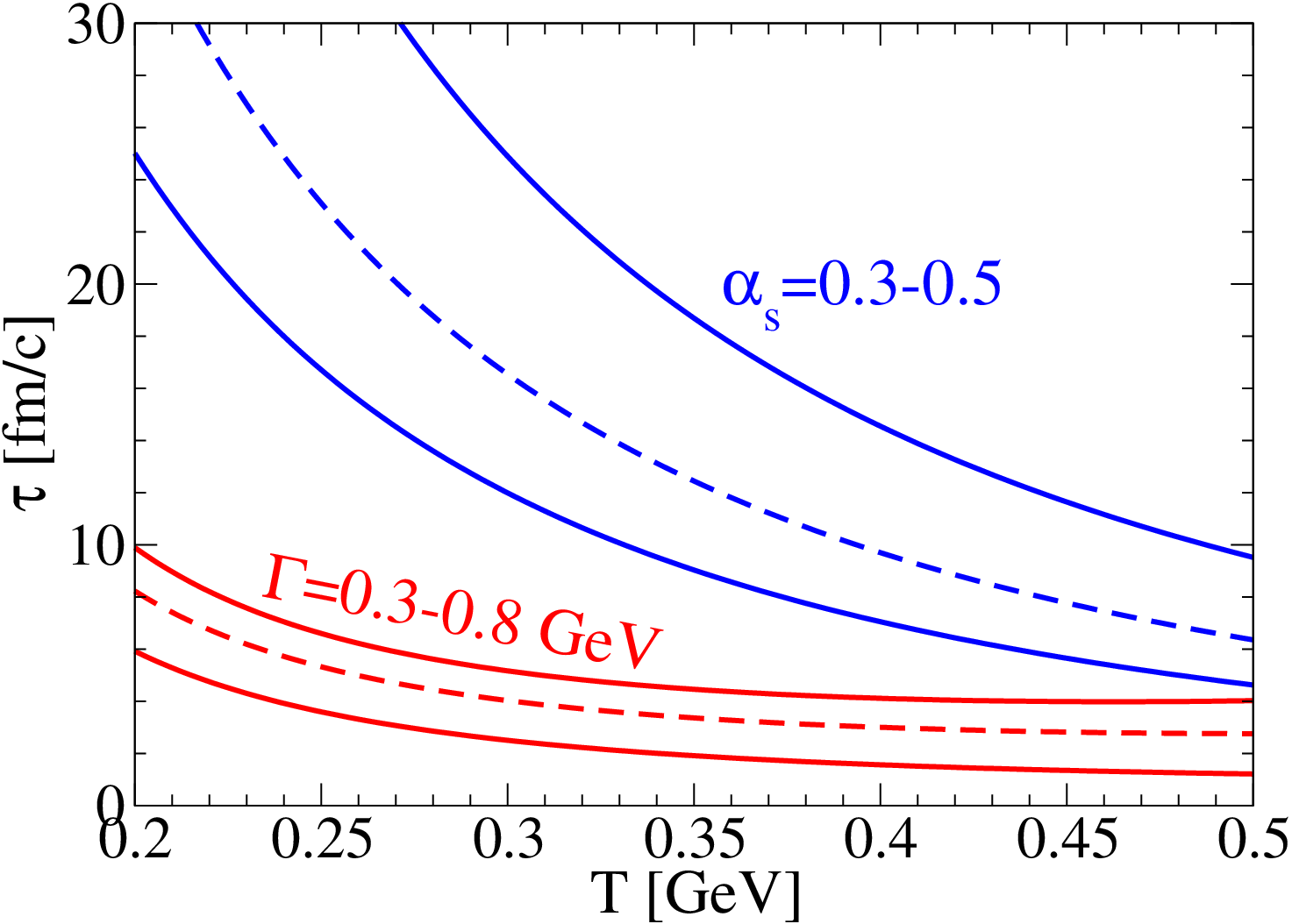}
}
                                                                                
\vspace{0.75cm}
                                                                                
\resizebox{0.45\textwidth}{!}{
\includegraphics{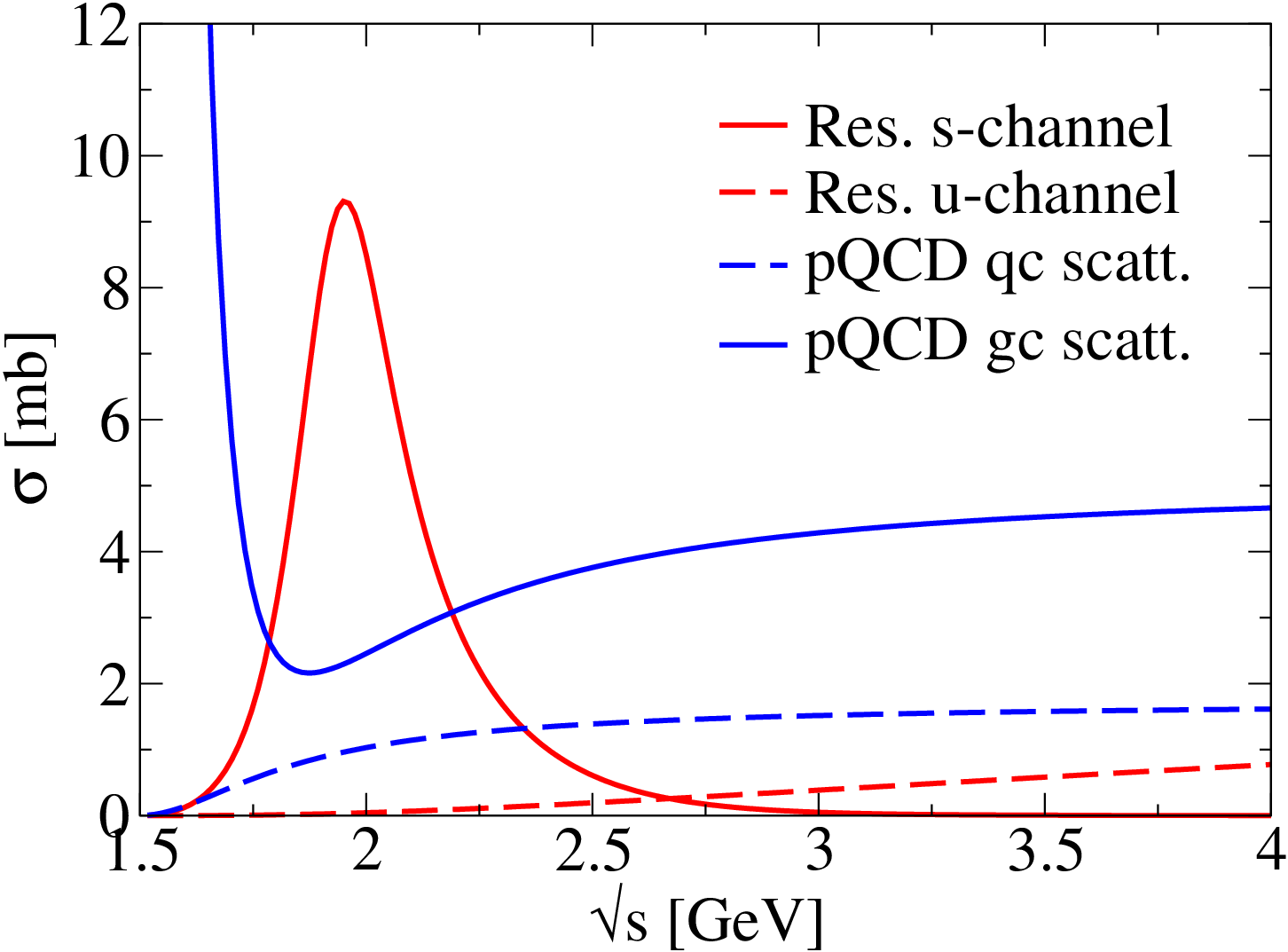}
}
\vspace{0.25cm}
\caption{Upper panel: $c$-quark equilibration times in QGP with
pQCD interactions (upper band; upper [lower] line corresponds
to $\alpha_s$=0.3[0.5])) and when adding
``$D$"-meson resonance rescattering (lower band; the range of resonance
widths indicates variations in the coupling constant of the
$c$-$q$-$D$ vertex with the upper [lower] line corresponding
to $\Gamma_D$=0.3[0.8]GeV)~\cite{HR05}. Lower panel: 
underlying total $c$-parton cross sections.}
\label{fig_cquark}
\end{figure}
Concerning (ii), it has been known for a while~\cite{Svet88} (and 
confirmed in Refs.~\cite{Must97,HR05}) that
perturbative $c$-quark rescattering off quarks and gluons in the QGP
implies kinetic relaxation times $\tau_c^{therm}$$\gsim$10fm/c for
$T$$\simeq$400MeV, too long to achieve thermalization at RHIC.
However, as shown recently~\cite{HR05}, nonperturbative 
rescattering in the QGP can lead to a substantial acceleration of 
equilibration: implementing the notion of $D$-meson-like resonances
within a Fokker-Planck equation, 
a reduction of $\tau_c^{therm}$ by a factor 
of $\sim$3 as compared to using pQCD cross sections has been found 
(for $T$$\le$2$T_c$), cf.~upper panel of Fig.~\ref{fig_cquark}
(similar for $b$-quarks, but with absolute values
$\tau^{therm}_b$$\simeq$4$\tau_c^{therm}$). 
The main difference in the two mechanisms resides not so much in the 
total cross sections (lower panel of Fig.~\ref{fig_cquark}), but in 
the isotropic angular distribution for the resonance case as opposed to
forward-dominated pQCD scattering. It has also
been noted~\cite{HR05} that the efficiency of this mechanism relies 
to a significant part on the $D$-states being located {\em above} the
$c$-$q$ threshold (\ie, not being boundstates, which renders them
inaccessible in 2$\to$2 scattering, especially due to the 
thermal energies carried by the light quarks).
It will be very valuable to check this in QCD lattice calculations, 
as well as whether previously found $q$-$\bar q$ and $Q$-$\bar Q$ 
states carry over to the $Q$-$\bar q$ sector. 
Furthermore, an increasing population of (colorless) ``hadronic"
states in the cooling process toward $T_c$ could serve as a mechanism
to put phenomenologically successful coalescence models on a 
firmer basis (also in the light-quark sector).
The in-medium mass of open-charm states in the QGP also bears
on the production of charmonia, as will be seen in the following
section.

From a phenomenological point of view, it should be kept in mind that 
any process contributing to elastic $c$-quark scattering in the QGP, 
$c+X_1\to c+X_2$, in principle also gives rise to secondary $c\bar c$ 
production in the crossed channel, $X_1+\bar X_2\to c+\bar c$, 
which can be constrained experimentally by total $c\bar c$ yields 
(including nontrivial centrality dependencies).
{\it E.g.}, in Ref.~\cite{Mol05} it has been
pointed out that when upscaling the perturbative $gc\to gc$ cross
section by a factor of 3 (to generate an elliptic flow comparable to light
quarks), secondary charm production is at the 40-50\% level of
the primordial yield in central $Au$-$Au$($\sqrt{s}$=200AGeV). 
This is expected to be less pronounced for heavier exchange particles, 
such as ``$D$"-mesons.  

\section{Charmonium in the QGP}
\label{sec_onium}
\begin{figure}[!t]
\vspace{0.5cm}
\centering
\resizebox{0.42\textwidth}{!}{
\includegraphics{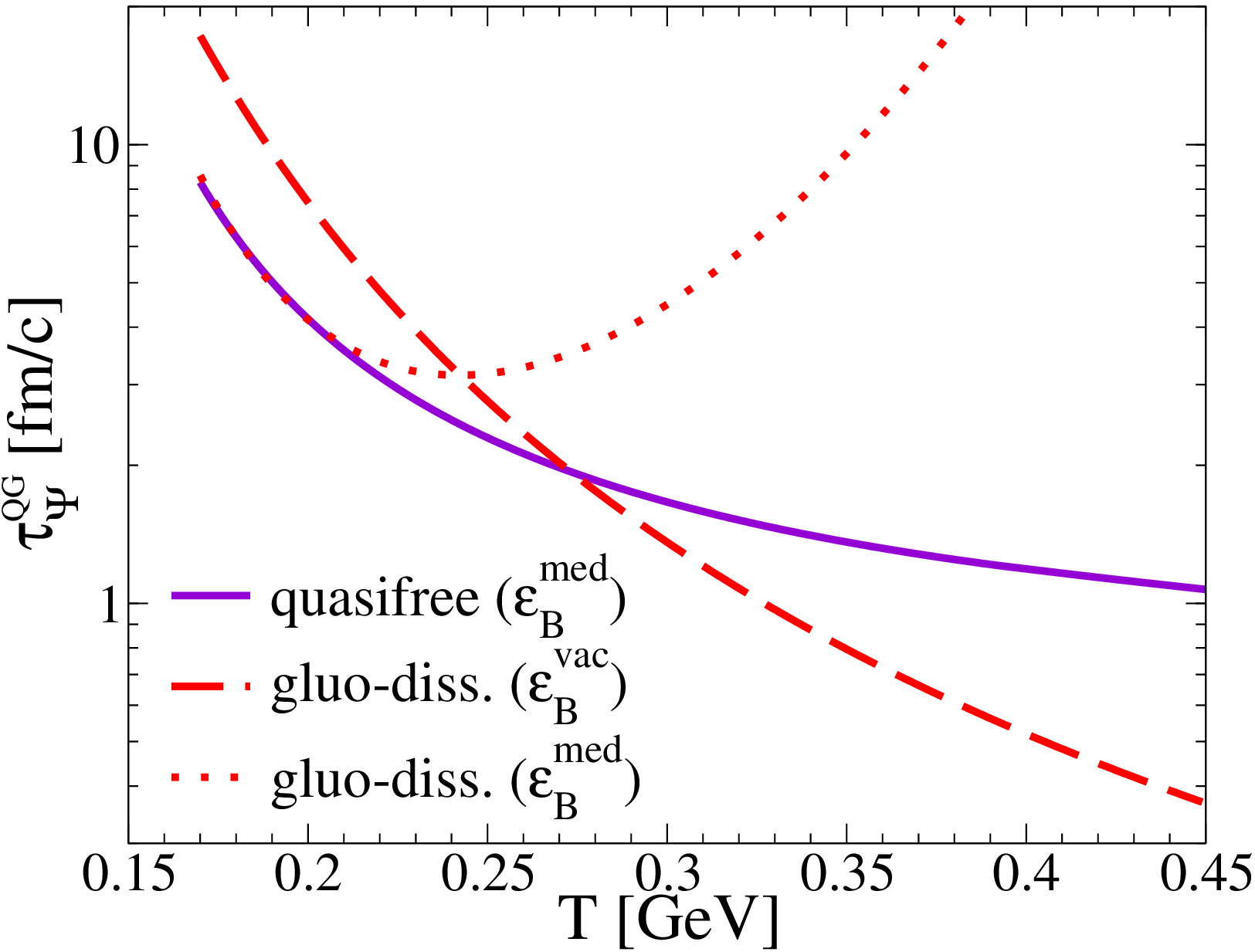}
}

\vspace{1.0cm}

\resizebox{0.46\textwidth}{!}{
\includegraphics{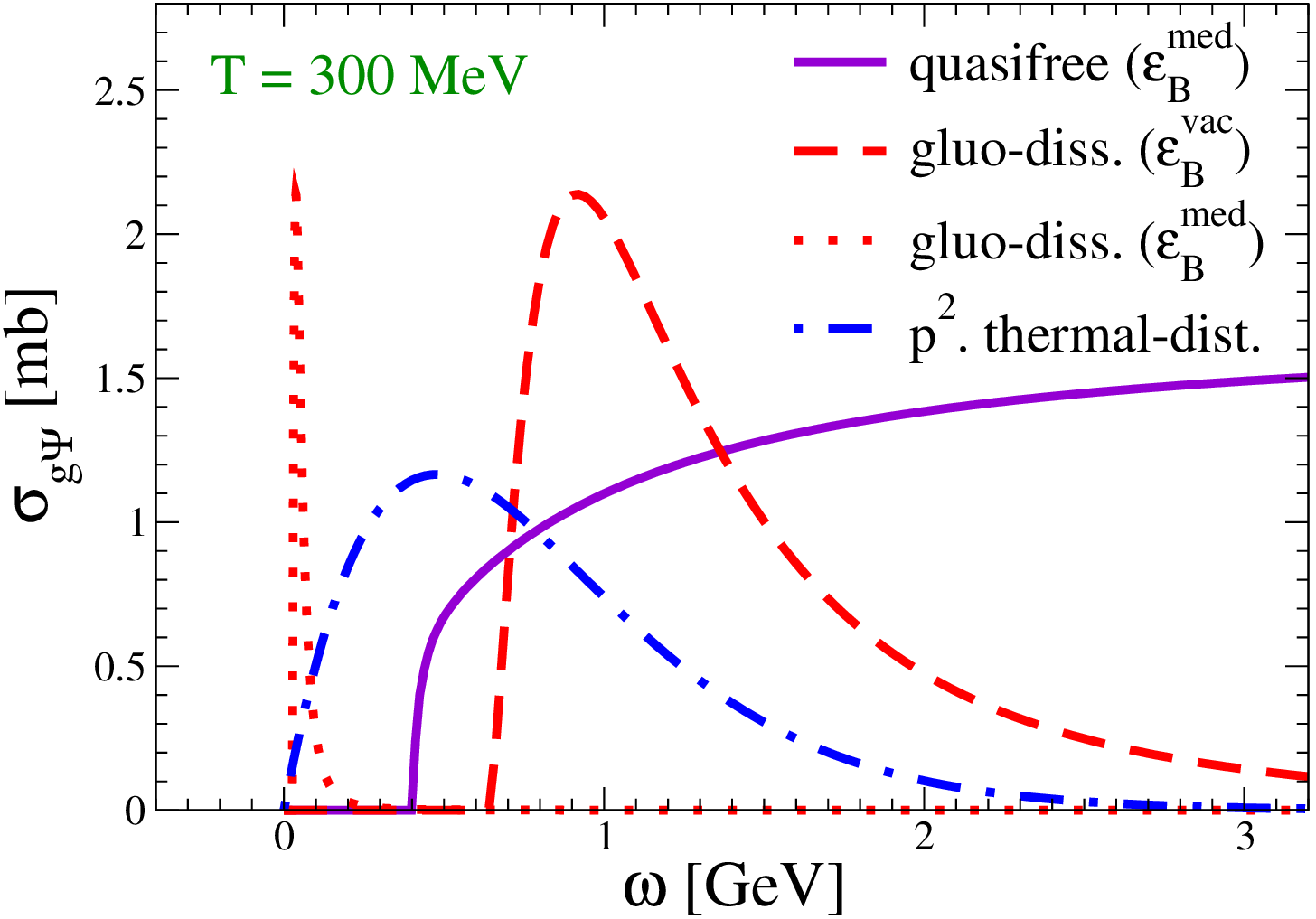}
}
\vspace{0.3cm}
\caption{Upper panel: $J/\psi$ lifetimes in the QGP
using gluodissociation~\protect\cite{BP79} with
vacuum (dashed line) and in-medium reduced (dotted line) 
binding energy, as well as quasifree dissociation~\protect\cite{GR01}
with in-medium reduced binding energy (solid line).
Lower panel: pertinent cross sections (line identification
as in upper panel) relative to thermal parton distribution 
functions (dash-dotted line).}
\label{fig_psi-xsec}
\end{figure}
A central quantity in evaluating medium effects on quarkonium
states, $\Psi$, in a heavy-ion collision are their inelastic cross
sections, $\sigma_\Psi^{diss}$, with partons in the QGP, determining
the pertinent dissociation rate as
\beq
\Gamma_\Psi = (\tau_\Psi)^{-1} = \int \frac{d^3k}{(2\pi)^3} \ f^{q,g}(\omega_k,T)  \
v_{rel} \ \sigma^{diss}_\Psi(s) \ .
\eeq
A widely used model for $\sigma^{diss}_\Psi$ is the gluon-absorption
break-up~\cite{Shu78,BP79}, $g+\Psi\to c+\bar c$, characterized by a
pronounced maximum at a gluon energy
$\omega_{max}$$\simeq$1.5$\epsilon_B$ ($\epsilon_B$: quarkonium binding
energy), see lower panel in Fig.~\ref{fig_psi-xsec}.
For $J/\psi$ mesons with their free binding energy, 
$\epsilon_B^{vac}$=640MeV, $\omega_{max}$
essentially coincides with thermal gluon energies,
$\omega$=3$T$, for $T$$\simeq$300MeV.
Debye screening of the $Q$-$\bar Q$ potential in the QGP is, however,
expected to substantially reduce $\epsilon_B$~\cite{KMS86}. This renders
gluodissociation an increasingly inefficient process at higher $T$
due to a shrinking break-up kinematics, cf.~dotted lines in
Fig.~\ref{fig_psi-xsec}.
For small $\epsilon_B$,  "quasifree" dissociation~\cite{GR01,RG04},
$g(q,\bar q) +\Psi\to c+\bar c + g(q,\bar q)$, 
albeit formally suppressed by one power of
$\alpha_s$, has been identified as a more important mechanism due to
much larger overlap with the thermal (quark + gluon) phase space 
(cf.~solid and dash-dotted lines in the lower panel of 
Fig.~\ref{fig_psi-xsec}).

\begin{figure}[!t]
\vspace{0.7cm}
\centering
\resizebox{0.45\textwidth}{!}{
\includegraphics{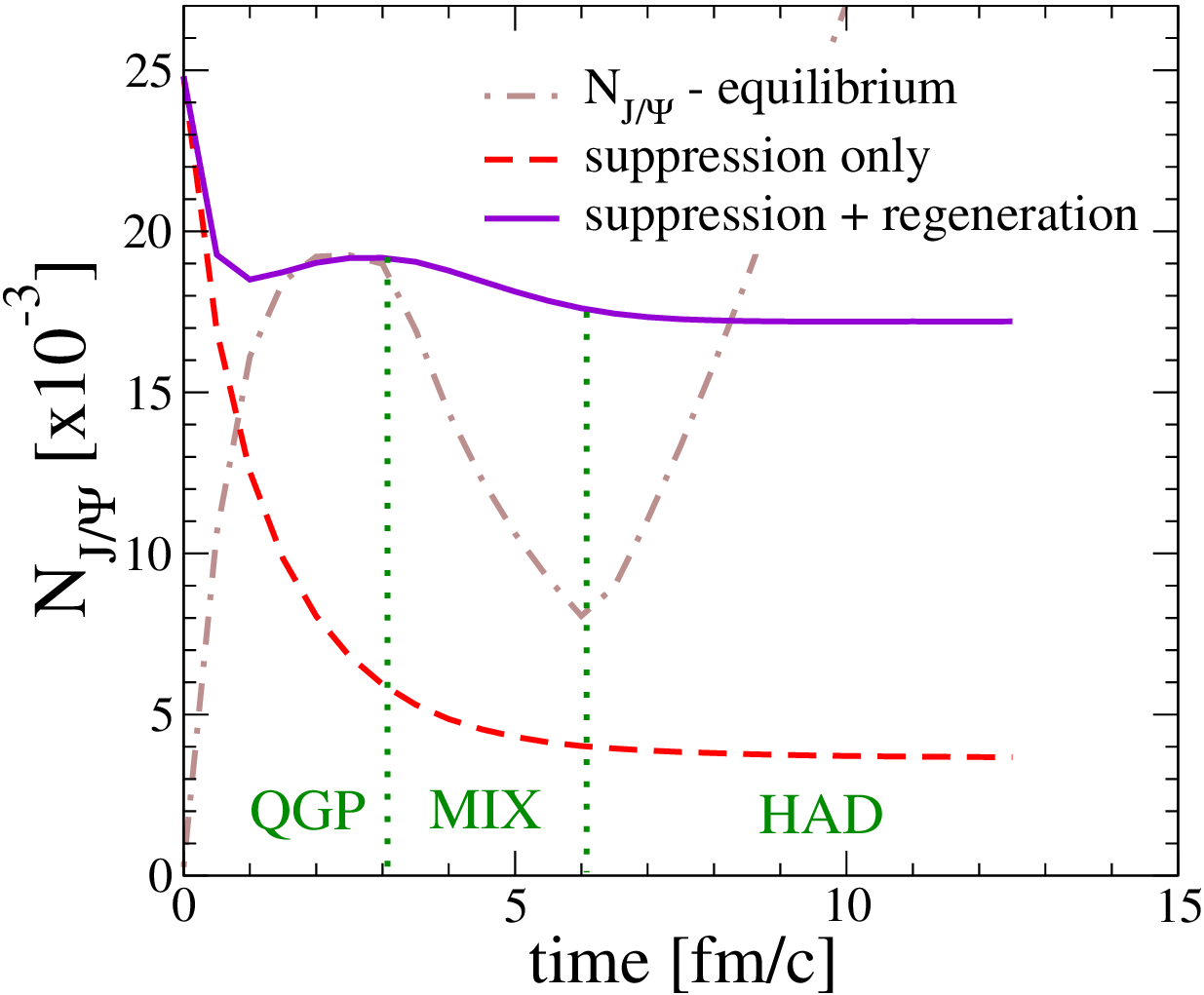}
}
\vspace{0.3cm}
\caption{Time evolution of the $J/\psi$ abundance in
central $Au$-$Au$($\sqrt{s_{NN}}$=200GeV) based on a solution
of the rate Eq.~(\protect\ref{rate-eq}) in an expanding thermal fireball
including in-medium effects on
both open and hidden charm mesons~\protect\cite{GRB04}.
Dashed line: suppression only (no gain term); dash-dotted line:
temperature-dependent equilibrium number, $N_{J/\psi}^{eq}(T;\gamma_c)$;
solid line: total number $N_{J/\psi}(t)$.}
\label{fig_time-evo}
\end{figure}
If the number of heavy quarks in a heavy-ion collision is
large enough, their recombination into quarkonia could become a
significant (or even dominant) contribution to the final
yield~\cite{pbm00,Thews01,Goren01,GR01,Zhang02,Brat03}.
The conditions for this to happen can be assessed in terms of
a simple rate equation for the time evolution of the number of
$\Psi$'s,
\beq
\frac{dN_\Psi}{dt} = - \Gamma_\Psi \left( N_\Psi -N_\Psi^{eq}\right)
\ .
\label{rate-eq}
\eeq
Besides the reaction rate $\Gamma_\Psi$, the other quantity governing
the evolution of $N_\Psi$ is the equilibrium abundance,
$N^{eq}_\Psi(T;\gamma_c)$, which determines $\Psi$ regeneration, \ie,
the gain term in Eq.~(\ref{rate-eq}), as required by detailed balance.
$N_\Psi^{eq}(T;\gamma_c)$ is typically evaluated in the canonical
ensemble with the total number of (primordial) $c\bar c$ pairs fixed
via a fugacity $\gamma_c$=$\gamma_{\bar c}$=e$^{\mu_c/T}$.
This implies that $N_\Psi^{eq}(T;\gamma_c)$ is sensitive to the 
open-charm spectrum, in particular (in-medium) masses of $c$-quarks 
(or $D$-mesons)~\cite{RG04,GRB04}; \eg, if $m_c^*$
(or $m_D^*$) is reduced in matter (with $m_\Psi$ constant), $c$- and
$\bar c$-quarks are thermally favored to occur in open-charm states, 
thus reducing $N_\Psi^{eq}(T;\gamma_c)$.
Finally, the gain term depends on the $c$-quark momentum distributions;
its particularly simple form in Eq.~(\ref{rate-eq}), based on
thermalized $c$-quarks, illustrates the impact of $c$-quark
rescattering (as discussed in the previous section) on charmonia.
Therefore, thermalization of $c$-quarks opens the window on equilibrium
properties of both open and hidden charm, \ie, their masses encoded
in $N_\Psi^{eq}(T;\gamma_c)$, as well as charmonium widths ($\Gamma_\Psi$).
A sensitive observable to distinguish direct and regenerated $J/\psi$'s
turns out to be their elliptic flow,
$v_2^\Psi$~\cite{Muell04,Wang02,Greco04,Brat05}. If only suppression is
operative, $v_2^\Psi$ reaches a maximal value of $\sim$2-3\%~\cite{Wang02}, 
while it grows up to $\sim$15\% at $p_t^\Psi$$\simeq$4GeV for thermal 
$c$-$\bar c$ coalescence~\cite{Greco04}.

\begin{figure}[!t]
\vspace{0.6cm}
\centering
\resizebox{0.47\textwidth}{!}{
\includegraphics{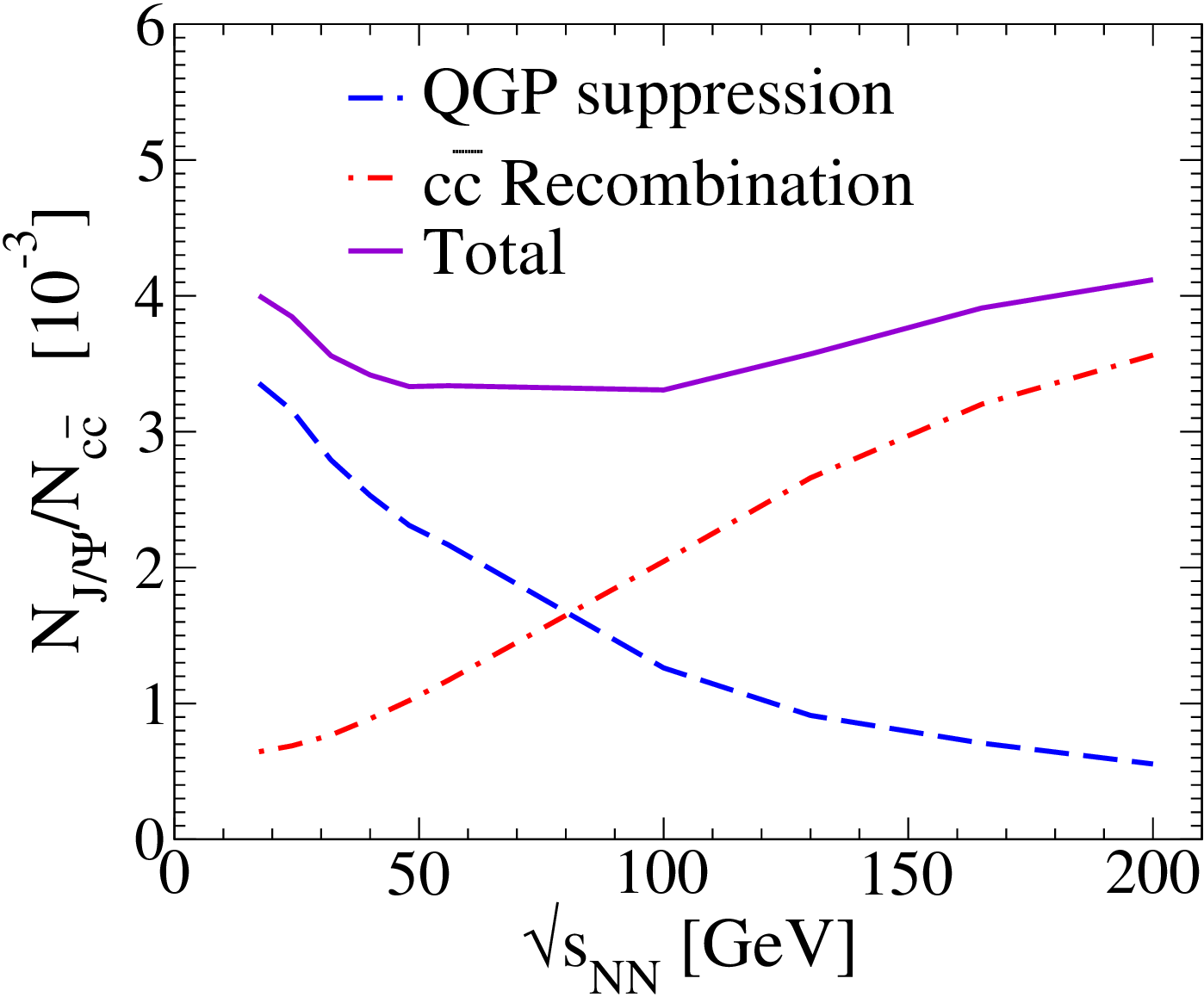}
}
\vspace{0.1cm}
\caption{Theoretical predictions for the excitation function of
$J/\psi$ production in central $Au$-$Au$ collisions. Suppression
(dashed line) and regeneration (dash-dotted line) components combine
into a rather flat energy dependence for the total yield
(solid line)~\protect\cite{GR01}.}
\label{fig_excit}
\end{figure}
A calculation~\cite{GRB04} of the time evolution of $N_{J/\psi}$ in 
central $Au$-$Au$($\sqrt{s}$=200AGeV) 
based on Eq.~(\ref{rate-eq}) including in-medium masses of open-charm 
and reduced $J/\Psi$ binding energies, as well as incomplete
thermalization of $c$-quarks in the early stages, is displayed in
Fig.~\ref{fig_time-evo}. One finds that the $J/\psi$ yield equilibrates 
close to the phase boundary, with the major contribution arising
from regeneration in the QGP and little changes in the ``mixed" and 
hadronic phase. Note that this result crucially hinges on the notion 
of the $J/\psi$ surviving as a resonance in the QGP under RHIC 
conditions, $T$$\le$2$T_c$.  
\begin{figure}[!t]
\vspace{0.7cm}
\resizebox{0.45\textwidth}{!}{
\includegraphics{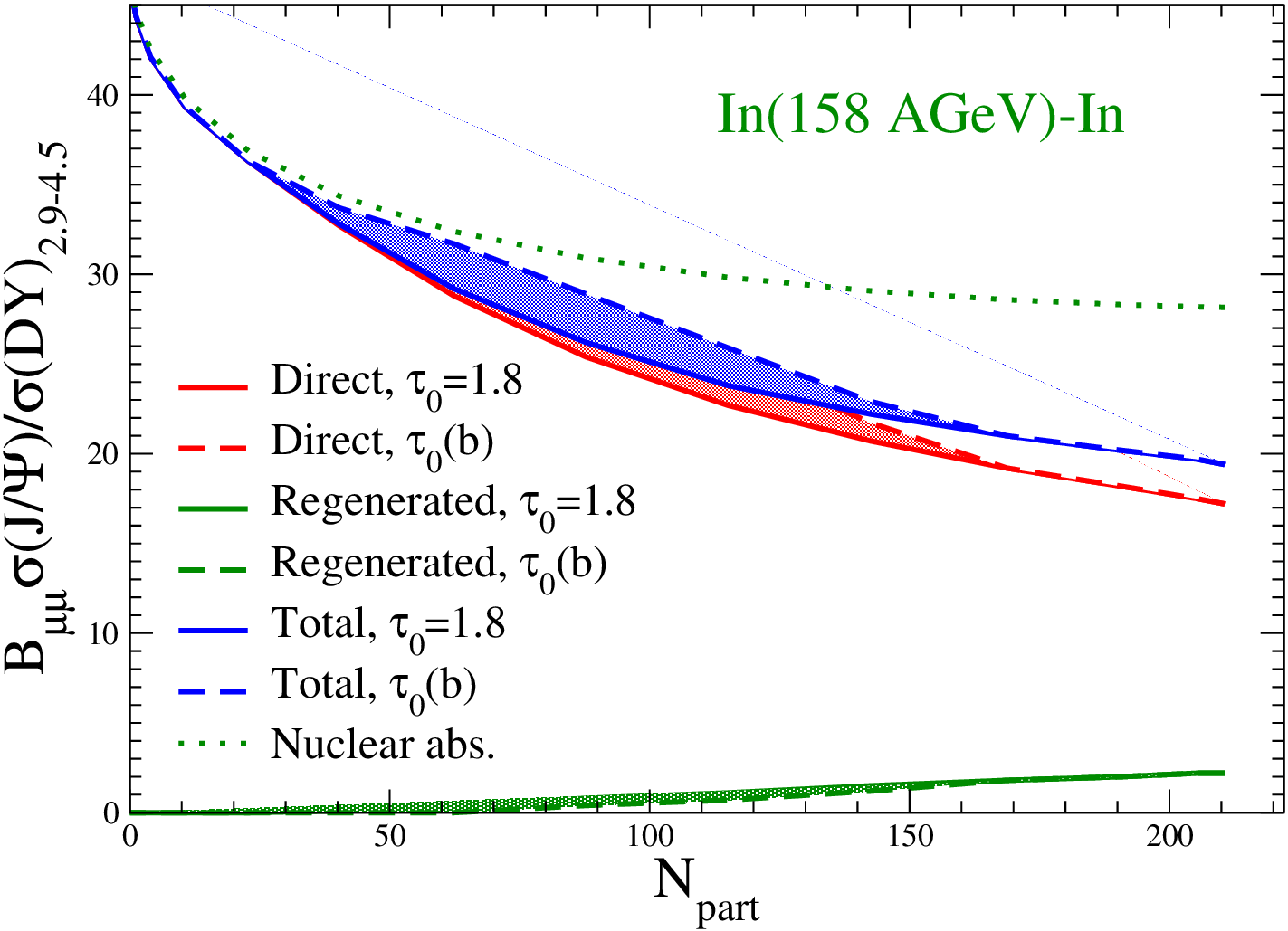}
}

\vspace{1.05cm}

\resizebox{0.45\textwidth}{!}{
\includegraphics{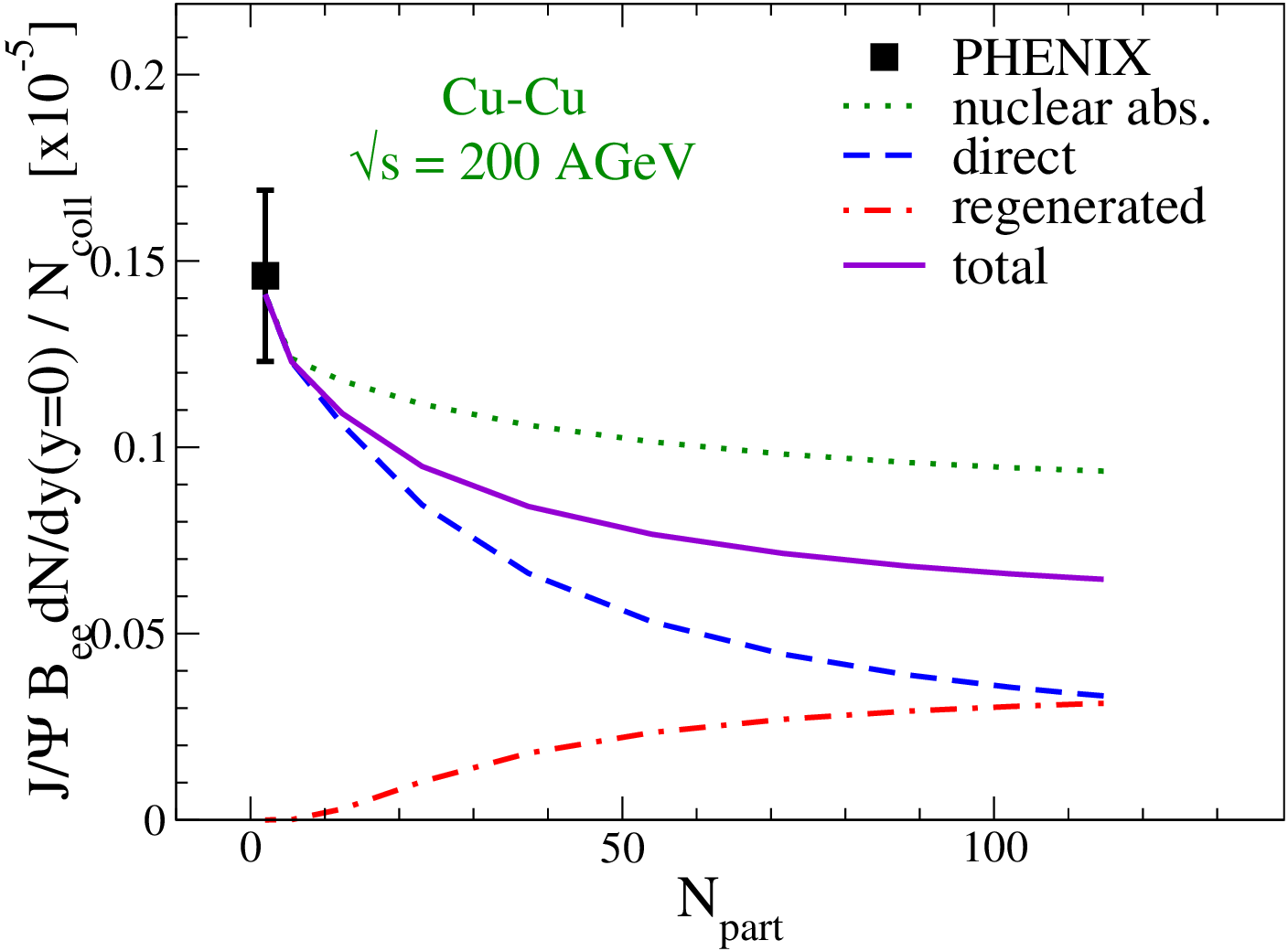}
}
\vspace{0.3cm}
\caption{Theoretical predictions for the centrality dependence of 
$J/\psi$ production in intermediate-size-ion collisions at SPS (upper 
panel) and RHIC (lower panel), 
including both suppression and regeneration processes~\cite{GRB04}.
The bands in the upper panel reflect uncertainties in the
formation time which is expected to increase at lower collision
centrality (dashed lines).}
\label{fig_imsize}
\end{figure}
The final yield is a factor of $\sim$4-5 increased over
a scenario with suppression only.
The situation is quite different at SPS energies ($\sqrt{s}$=17.3AGeV):
with an expected open-charm number $N_{c\bar c}$$\simeq$0.2 in central
$Pb$-$Pb$, secondary charmonium formation is negligible and $J/\psi$
{\em suppression} is the main mechanism at work.
Obviously, this calls for mapping out the excitation function for
$\sqrt{s}$=20-200GeV (accessible at RHIC), as suggested in
Ref.~\cite{GR01}. Based on Fig.~\ref{fig_excit} one expects a
transition from a suppression-dominated regime (SPS or low RHIC
energies) to a regeneration-dominated one at
$\sqrt{s}$$\gsim$100AGeV, resulting in a rather flat energy dependence
(possibly with a shallow minimum).

Complementary information on the interplay between primordial and
secondary $J/\psi$ production can be extracted by going to smaller
nuclear collision systems.
Pertinent predictions are shown in Fig.~\ref{fig_imsize}, reconfirming
the absence of noticeable regeneration at SPS (as well as a smooth
centrality dependence; upper panel), but an approximately equal amount of 
primordial and regenerated $J/\psi$'s for central $Cu$-$Cu$ at RHIC
(lower panel).

\section{Conclusions}
\label{sec_concl}
Hadrons containing heavy quarks are excellent probes of the environment
formed in nuclear reactions. 
Evidence for coalescence mechanisms in elementary hadronic reactions
finds its natural extension for both $D$-mesons and charmonia
to heavy-ion collisions. In addition, at RHIC, the produced medium
appears to interact strongly enough to thermalize $c$-quarks (but
not $b$-quarks).
If confirmed, ``$D$"-meson resonance formation in the QGP (coupled
with pertinent coalescence at $T_c$) might be the key to a 
simultaneous understanding of (suppressed) $p_t$ spectra and (large) 
elliptic flow of $D$-mesons (and single electrons). The transition 
into a perturbative energy-loss picture could be shif\-ted to higher 
$p_t$ than for light hadrons.
Resonance states in the QGP also have substantial impact on charmonium
production, facilitating their regeneration in the 1-2$T_c$ re\-gime 
where inelastic collision rates are high. Here, thermalization of 
$c$-quarks would enable a rather direct window on spectral properties 
of open and hidden charm, \ie, their masses and widths. 
Work in progress on $\Upsilon$ production~\cite{Lum05} seems to 
indicate, however,
that even at LHC their suppression is prevalent, due to a lack of
thermalization of bottom quarks. Thus, a simultaneous observation
of $\Upsilon$ suppression and the absence thereof for $J/\psi$
at collider energies would provide strong evidence for secondary
charmonium production.

Among the main challenges yet to be met is establishing connections 
of heavy-quark observables to (``pseudo"-) order parameters of
the QCD phase transition~\cite{NS98}. With low-lying charmonia 
possibly surviving 
up to 2$T_c$, their dissolution evades a direct relation to $T_c$.
A suitable quantity could be their inelastic width, which in model
calculations is typically quite different (smaller) in 
the hadronic compared to the QGP phase~\cite{RG04}. 
Quenched lattice calculations~\cite{Umeda05}
indicate a similar trend, but unquenching has to be awaited
for more definite conclusions. We also mention the recent
work of Ref.~\cite{Terra05}, where an increase in transverse-momentum 
fluctuations of open-charm states has been linked to a first-order 
transition.  

Looking into the future, it seems that the combined experimental and 
theoretical analysis of heavy-quark observables in ultrarelativistic 
heavy-ion collisions is on a promising path toward providing a 
milestone in the identification of the QGP. 
\\

{\bf Acknowledgment.} I thank the organizers
for the invitation to a very informative conference, and L.~Grandchamp,
V.~Greco, H.~van~Hees, C.M.~Ko and E.V.~Shuryak for collaboration
on the presented topics.
This work has been supported in part by a U.S. National Science 
Foundation CAREER award under grant PHY-0449489.

\end{document}